\documentclass[conference]{IEEEtran}

\usepackage{cite}
\usepackage{amsmath,amssymb,amsfonts}
\usepackage{algorithm,algpseudocode}
\usepackage{url}            
\usepackage{booktabs}       
\usepackage{amsfonts}       
\usepackage{nicefrac}       
\usepackage{microtype}      
\usepackage{lipsum}
\usepackage{graphicx}
\usepackage{subcaption}
\usepackage{caption}
\usepackage{textcomp}
\usepackage{xcolor}
\usepackage{array}
\usepackage{tabularx}
\usepackage{multirow}
\usepackage{ragged2e}
\usepackage{xspace}
\usepackage{fancyhdr}

\newcommand{\defIntvl}{\mathcal{T}_d}
\newcommand{\refIntvl}{\mathcal{T}_r}
\newcommand{\Eav}{\ensuremath{\mathcal{E}}\xspace}
\newcommand{\Etx}{\ensuremath{E_{tx}}\xspace}
\newcommand{\lav}{\ensuremath{\Lambda}\xspace}
\newcommand{\lat}{\ensuremath{{L}}\xspace}

\fancypagestyle{copyrightstyle}{
  \fancyhf{} 
  \fancyfoot[C]{Copyright © 2026 goTenna Inc.} 
}

\def\BibTeX{{\rm B\kern-.05em{\sc i\kern-.025em b}\kern-.08em
    T\kern-.1667em\lower.7ex\hbox{E}\kern-.125emX}}

\begin{document}

\title{SQEEZ: Energy-Efficient Location Sharing in Mobile Ad Hoc Networks}

\author{\IEEEauthorblockN{
Ram Ramanathan\IEEEauthorrefmark{1}\IEEEauthorrefmark{2}, 
Dmitrii Dugaev\IEEEauthorrefmark{1},
Ryan Conyac\IEEEauthorrefmark{1},
Alon Mor\IEEEauthorrefmark{1},
Charlie Greenbacker\IEEEauthorrefmark{1}
}
\IEEEauthorblockA{\IEEEauthorrefmark{1}goTenna Inc., 101 Hudson Street, STE 1701, Jersey City, NJ 07302\\
\IEEEauthorblockA{\IEEEauthorrefmark{2}NLytica LLC, 82 Wendell Ave, STE 100, Pittsfield, MA 01201}
Email: ram@nlyticallc.com, dmitrii@gotenna.com, ryanc@gotenna.com, alon@gotenna.com, charlie@gotenna.com}
}

\maketitle
\thispagestyle{copyrightstyle}

\begin{abstract}
Periodic network-wide dissemination of node location data is crucial for shared situational awareness and collaborative mapping in mobile ad hoc and mesh networks for public safety, disaster relief, and military. A key challenge is to provide maximally accurate location information with minimal energy expenditure on part of the nodes. We present SQEEZ: a mechanism for reducing the Position Location Information (PLI) load that combines two orthogonal techniques: (1) adaptive suppression of location updates; and (2) temporal and inline compression of update packets. We describe the SQEEZ suppression and compression algorithms, analyze the tradeoff between location error and energy consumption, and introduce a new metric called \textit{Error-Penalized-Energy (EPE)} that normalizes the energy metric using the error incurred.
Our simulation results show that, in the range of parameters studied, SQEEZ improves the EPE-efficiency and scalability in a 30-node random waypoint scenario by up to 4.4x and 2.3x rexpectively; and increases the EPE-efficiency by 7.5x in a 9-node real-world network trace. Compression provides larger improvements than suppression at high mobilities and vice-versa at low mobilities.


\end{abstract}


\section{Introduction}
\label{sec:introduction}

In Mobile Ad Hoc Networks (MANETs) deployed for off-grid communications in scenarios such as disaster relief, wildland firefighting, and military operations, nodes periodically broadcast their locations over the MANET to provide shared situational awareness and collaborative mapping. A prime example of an application for this purpose is the Android Team Awareness Kit (ATAK) -- a geospatial infrastructure application that allows users to navigate using GPS and map data overlays, providing a real-time common operating picture. In a MANET environment, plugins for ATAK enable the periodic sharing of Position Location Information (PLI) messages, ensuring all team members have a consistent and up-to-date view of each others' locations.



The dissemination (broadcast) of PLI messages from every node to every other node imposes a significant load on a MANET, which increases with both increasing size and mobility.
With increasing size, more PLIs are generated and forwarded; and with increasing mobility, it is necessary to increase the location update frequency for accurate tracking. In order to provide long range and low SWaP (Size, Weight, and Power), many of the products supporting off-grid communications such as the goTenna Pro-X2~\cite{gotennaPro} end up having low bitrates of the order of 10-50 kbps, and run on batteries. In such systems, the increasing PLI load causes packet losses, drains device battery, and in general limits scalability.

We present SQEEZ (Suppression and Compression for Energy- and Error-Minimizing Location Sharing) for adaptively reducing the Position Location Information (PLI) packet load on the network to better scale with size and mobility. SQEEZ uses a combination of three techniques: (1) adaptive PLI \textit{suppression}, wherein the updates are selectively discarded at the originator if the node has not moved beyond a threshold; (2) \textit{temporal compression} wherein only the \textit{difference} between a PLI packet and a reference PLI packet is transmitted; and (3) a further \textit{inline compression} of this difference using the Dynamic Compact Control Language (DCCL). 

Whether or not SQEEZ is employed, there is an interesting tradeoff between the energy consumption, the packet delivery rate, and the location tracking accuracy as a function of the PLI update frequency. If the PLI update frequency is low, then the locations may not be sufficiently accurate, but if one increases the PLI frequency, the energy drain may be prohibitive. Moreover, if the load increases beyond a point, PLI packets may be dropped at the MANET network and MAC layers, further exacerbating location errors. We present a preliminary mathematical model to capture this tradeoff, and show that the product of location error and energy is a constant for a given velocity. 


SQEEZ has been designed to work with location sharing applications such as ATAK, in particular when these are deployed over a MANET. An example of such a deployment is over the goTenna Pro-X2 product~\cite{gotennaPro}, a very low SWaP-C (size, weight, power, and cost) mesh/MANET radio device.  
Widely used by military, law enforcement, and public safety personnel in the U.S. and around the world, the Pro-X2 utilizes goTenna's novel Aspen Grove protocol stack~\cite{Ramanathan19} that uses zero-control-packet routing protocols~\cite{Ramanathan18,ayushRamVine} to quickly and efficiently provide long-range short-burst mission-critical communications. 

Using a model of the goTenna Pro-X2 networking stack as the base and ATAK-based PLI message formats, we have simulated SQEEZ and evaluated the overall system performance on a variety of metrics. This includes metrics for energy consumption, location error, and packet delivery ratio -- plus a novel metric that we call \textit{Error-Penalized-Energy (EPE)} that captures the energy-accuracy tradeoff by "normalizing" the consumed energy with the incurred error. 
Our results show that SQEEZ improves the EPE-efficiency by up to 2x and 7x for random waypoint and real-world mobility models respectively, and increases scalability by up to 2.5x. Further, adaptive suppression is better than compression at low and medium mobilities, whereas compression outperforms suppression at higher mobilities. A key finding is that while decreasing PDRs increase location errors, but the increase starts at much lower PDRs than expected. 

The key contributions of this paper are as follows: (1) an architecture and algorithms for joint use of adaptive suppression and compression; (2) analytical expressions for location error and energy, and the tradeoff therein; (3) conception of a novel metric \textit{error-penalized-energy}; and (4) a comprehensive simulation study of SQEEZ, including each of its components, providing insights.

SQEEZ can be used to reduce the traffic load in any system or application that uses periodic broadcast messaging. 
For example, goTenna has developed a plugin for ATAK that offers support for various rates of PLI messaging based on operational needs.
Using SQEEZ within such a plugin, operators of off-grid mesh networks can significantly improve the network longevity and scalability with measurable and controllable bounds on location tracking errors.

\section{SQEEZ Architecture}
\label{sec:sqeez-architecture}

Currently, a typical system for providing situational awareness using location information dissemination -- we call this the \textit{baseline} system -- works as follows. An application such as ATAK~\cite{Usbeck2015} runs on each node. Using the in-built GPS, each node originates a Position Location Information (PLI) packet at periodic intervals containing the node's location (latitude/longitude) and other supporting information as the timestamp, sender identification, and expected location sharing frequency. This packet is then network-wide broadcast over an underlying MANET stack such as the goTenna Aspen Grove stack. Nodes receiving the packet update their map or other information structures using the received information from the sender.

The PLI period is a critical feature of such a system. The smaller it is (i.e., higher the PLI origination frequency), the better the tracking accuracy, but higher the load on the MANET. Below a certain PLI period, the load exceeds the network capacity and significant packet losses occur, leading to a \textit{decrease} in accuracy. 

The goal of SQEEZ is to reduce the PLI load in current systems using two orthogonal techniques:
\begin{itemize}
\item \textit{Suppression} of originated PLI packets if the distance between the current and previous locations is less than a threshold.
\item \textit{Compression} of PLI packets using two methods:

\begin{itemize}
\item \textit{Temporal compression}, wherein only the difference between an PLI and an "anchor" PLI is sent.
\item \textit{In-line compression} using the Dynamic Compact Control Language (DCCL), which efficiently encodes the packet into the smallest possible size
\end{itemize}

\end{itemize}

\begin{figure}[t]
\centering
\includegraphics[width=1.0\linewidth]{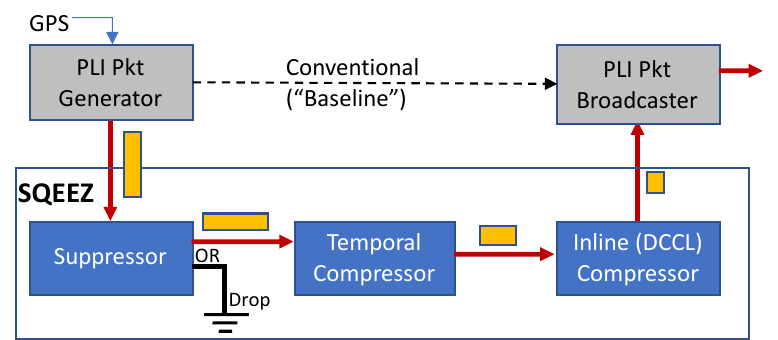}
\caption{The SQEEZ Architecture (source side).}
\label{fig:squeez-arch}
\vspace{-0.4cm}
\end{figure}

The architecture of SQEEZ is illustrated in Figure~\ref{fig:squeez-arch}. All operations are at the orginator and the receiver(s) of the PLI packet; the transit nodes in the MANET are not involved at all. SQEEZ is an augmentation of the baseline system wherein instead of sending an originated packet immediately, it is sent to SQEEZ instead. This architecture allows SQEEZ to coexist with the baseline system, requires minimal modifcation to it and facilitates flexible and incremental deployment. 

Within SQEEZ, the PLI packet first goes through the \textit{suppressor}, which decides whether to suppress the packet or not based on previously stored packets. If the packet is not suppressed, then it is given to the \textit{temporal compressor} which compresses the packet using a reference anchor. The packet is then further compressed by encoding via DCCL. At a receiver of the PLI, SQEEZ applies these operations in reverse -- first DCCL decoding, then temporal de-compressing, before handing the PLI to the application.

In many systems, the PLI is encrypted to prevent unauthorized access. If such encryption is present, it is applied at the very end of the SQEEZ process, i.e., after the DCCL compressor. A received packet is likewise decrypted before DCCL coding. Thus, SQEEZ is fully compatible with PLI encryption.

\section{SQEEZ Algorithms}
\label{sec:sqeez-algorithms}
In this section we describe each of the components depicted in Figure~\ref{fig:squeez-arch} in detail.

\subsection{Suppression}

A formal description of the SQEEZ Suppression algorithm is given in Algorithm \ref{alg:suppression}, and explained below.

The Suppressor inherits the \textit{default PLI interval} $\defIntvl$ from the baseline, and uses two additional parameters: the \textit{refresh interval} $\refIntvl$, which is the time after which a PLI is transmitted even if there has been no change in the location; and the \textit{threshold distance} $\Delta$, which is the radius within which PLIs are suppressed.

The SQEEZ Suppressor uses two states for the inferred status of the node on which it runs: \textit{mobile} and \textit{stationary}. The initial state is \textit{mobile}. The state is inspected every default period $\defIntvl$ and transitions from \textit{mobile} to \textit{stationary} if the node is within $\Delta$ of the previous update's location. Specifically, the SQEEZ suppressor reads the current location information $L_c$ from the PLI, and compares it with the stored location $L_s$ of the most recent previous PLI that was actually shared. If the Euclidean distance $d(L_c, L_s)$ is less than $\Delta$, the packet is dropped (suppressed), unless the time elapsed since a previous transmission exceeds $\refIntvl$, in which case the packet is sent. 

Conversely, the state transitions from \textit{stationary} to \textit{mobile} if  $d(L_c, L_s)$ $>=$ $\Delta$. In the \textit{mobile} state, a PLI is never suppressed. 

The default PLI interval is often a part of the PLI packet, and may also be called \textit{location sharing frequency}. This serves to inform a receiver when to expect the next packet, based on which the receiver can mark a sender ``unavailable" if no PLI was received from the node after some multiple of the default PLI interval. Accordingly, the SQEEZ Suppressor sets this field to  $\defIntvl$ when a PLI is sent in the \textit{mobile} state, and to $\refIntvl$ when in \textit{stationary} state. The latter indicates to the receiver not to expect a PLI for another $\refIntvl$ time even though the default interval may be less. Further, an additional PLI is sent when transitioning from the mobile to stationary state whose sole purpose is to inform the receivers of the new location sharing frequency $\refIntvl$ (see lines 11-14 in Algorithm~\ref{alg:suppression}) to update their expectations for the next PLI.

As per Algorithm \ref{alg:suppression} updates are only sent at the first expiration of the $\defIntvl$ timer after the $\Delta$-threshold was crossed. We refer to this as the \textit{discrete} version of suppression. In contrast, we could have a \textit{continuous} version where the location is constantly sampled, and an update is sent as soon as the $\Delta$-threshold is crossed. This can be thought of as a special case of Algorithm \ref{alg:suppression} when $\defIntvl$ is equal to the sampling interval. The continuous version has lower location error but uses more energy and originates more packets. In the remainder of the paper, all references to Suppression assume the discrete version unless explicitly specified otherwise. 

\begin{algorithm}
\caption{Suppression Algorithm}
\label{alg:suppression}
\begin{algorithmic}[1]
\State \textit{Input}: Generated PLI packet (curPLI)
\State \textit{Configured Parameters}: $\defIntvl$, $\refIntvl$, $\Delta$ \Comment See text
\State \textit{Local Store}: Most recently sent PLI (lastSentPLI)
\State   
\Procedure{everyDefaultInterval}{$\defIntvl$, curPLI}
\If{DISTANCE(curPLI.loc, lastSentPLI.loc) $\geq$ $\Delta$}
\State state $\gets$ \textit{MOBILE}
\State curPLI.locSharingFreq $\gets$ $\defIntvl$
\State network-broadcast curPLI 
\Else
\If{state is \textit{MOBILE}}
\State state $\gets$ \textit{STATIONARY}
\State curPLI.locSharingFreq $\gets$ $\refIntvl$
\State network-broadcast curPLI
\Else
\State discard curPLI
\EndIf
\EndIf
\EndProcedure
\State  
\Procedure{everyRefreshInterval}{$\refIntvl$, curPLI}
\State network-broadcast curPLI unless done within last $\refIntvl$
\EndProcedure
\end{algorithmic}
\end{algorithm}

\subsection{Temporal Compression}

Position Location Information (PLI) packets typically have significant redundancy along the ``temporal axis." For instance, consecutive PLI messages often differ only in a few specific fields (e.g., timestamp, coordinates), while others remain constant (e.g., callsign, team color). Temporal compression exploits this redundancy using a simple idea: send the unchanging fields once and send only the difference in subsequent packets. While the basic idea is well-known and has been used in solutions such as \textit{Robust Header Compression (ROHC)} for IP networks, the design of a particular temporal compression solution depends heavily on the packet fields, which are very different for PLIs compared to IP headers. A particular challenge is robustness against packet losses while retaining efficiency: this requires proper classification of fields in terms of how often they change.

We present below the design in the context of the goTenna plugin for ATAK. Table~\ref{tab:pli_fields} enumerates the fields within a typical PLI protobuf used in the goTenna application, along with their persistence and classification. We classify fields into three categories: STATIC (rarely, if ever, changes), SEMI-STATIC (changes slowly), and CHANGING (frequently changes).

\begin{table}[ht] 
\centering 
\begin{tabular}{|l|c|l|l|} \hline 
\textbf{Field} & \textbf{Bytes} & \textbf{Class} & \textbf{Persistence} \\ \hline
Timestamp & 8 & CHANGING & Secs - Mins \\
Message Type & 1 & STATIC & N/A \\
App Code & 1-8 & SEMI-STATIC & Days - Months \\ 
Sender GID & 8 & STATIC & N/A \\ 
Encryption ID & $\sim$3 & SEMI-STATIC & Days - Months \\
Init. Vector & $\sim$12 & CHANGING & Secs - Mins \\
Coordinate & 11-16 & SEMI-STATIC & Secs - Years \\ 
Loc. Share Freq & 4 & SEMI-STATIC & Hours - Days \\ 
TAK How & 1 & STATIC & N/A \\ 
Loc. Accuracy & 4 & STATIC & N/A \\ 
Loc. Type & 1 & STATIC & N/A \\ 
Team Color & 1 & STATIC & N/A \\ 
Sender UUID & 8-16 & STATIC & N/A \\ 
Sender Callsign & Var. & SEMI-STATIC & Days - Months \\ \hline 
\end{tabular} 
\caption{\bf \small PLI Fields. Class assigned by SQEEZ} 
\label{tab:pli_fields} 
\end{table}

As apparent from Table~\ref{tab:pli_fields}, a significant portion of the packet consists of STATIC or SEMI-STATIC fields (such as Sender GID and Team Color) which do not need to be re-transmitted in every update.

To manage the retrieval of these fields at the receiver, we utilize \textit{anchors} as keys for the shared context. We employ the Sender-UUID as the anchor for STATIC fields, as this is assigned to a user device and persists for the duration of the mission. For SEMI-STATIC fields, we introduce a new 1-byte Context-ID. The reason for a separate Context-ID in addition to the Sender-UUID is as follows: relying solely on UUID would require re-sending all semi-static fields in every message to be resilient to packet loss.\footnote{If a semi-static field changes, and the packet with the changed field gets lost, retrieving using the UUID anchor alone could result in the wrong value.} Using only a Context-ID would necessitate sending static fields whenever a semi-static field changes. Our dual-anchor approach balances robustness and efficiency.

The sender algorithm is given in Algorithm~\ref{alg:temporal-compression-sender} and functions as follows. Upon receiving a PLI from the application, the sender compares the STATIC fields against the stored context for the UUID. If there is a mismatch, a new context is generated, and an uncompressed packet is sent. If the STATIC fields match, the sender then checks the SEMI-STATIC fields against the stored context-id-anchored semi-static context. A mismatch here triggers the generation of a new Context-ID, and a ``semi-compressed" packet (containing UUID, new Context-ID, SEMI-STATIC, and CHANGING fields) is transmitted. If both match, a fully compressed packet containing only the Context-ID and CHANGING fields is sent. Periodically (including at the very beginning), the entire PLI with a new Context-id is sent as a \textit{soft-state refresh} to account for unforeseen anchor losses.

\begin{algorithm}
\caption{Temporal Compression: \bf{Sender Side}}
\label{alg:temporal-compression-sender}
\begin{algorithmic}[1]
\State \textit{Input}: Generated PLI packet (PLI)
\State \textit{Output}: Compressed PLI packet (cPLI)
\State \textit{Local Store}: Indexed Contexts
\State   
$U$ $\gets$ PLI.sender-UUID
\If{PLI.static $\neq$ stored-context[$U$]}
\State stored-context[$U$] $\gets$ PLI.static 
\State cPLI $\gets$ PLI \Comment Uncompressed
\Else
\If{PLI.(static+semi-static) $\neq$ stored-context[any]}
\State Generate new Context-ID $C$
\State stored-context[C] $\gets$ PLI.(static+semi-static)
\State Create semi-compressed cPLI with fields: 
\State \;\; UUID=$U$, ContextID=$C$, semi-static, changing
\Else
\State Retrieve match's Context-ID $C$
\State Create compressed cPLI with fields:
\State \;\; Context-ID = $C$, changing
\EndIf
\EndIf
\State return cPLI
\end{algorithmic}
\end{algorithm}

\begin{algorithm}
\caption{Temporal Compression: \bf{Receiver Side}}
\label{alg:temporal-compression-receiver}
\begin{algorithmic}[1]
\State \textit{Input}: (Possibly) Compressed PLI packet (cPLI)
\State \textit{Output}: Original (uncompressed) PLI packet (PLI)
\State \textit{Local Store}: Indexed Contexts
\State   
$U$ $\gets$ cPLI.sender-UUID
\If{cPLI is Uncompressed}   \Comment All fields present
\State stored[$U$] $\gets$ cPLI.static
\State PLI $\gets$ cPLI
\Else
\If{cPLI is Semi-compressed} \Comment No Static fields
\State Retrieve static $\gets$ Stored[U] 
\State PLI $\gets$ static + cPLI.semi-static + cPLI.changing
\State stored[cPLI.ContextID] $\gets$ static + cPLI.semi-static
\Else \Comment \textit{is fully compressed}
\State Retrieve static+semi $\gets$ stored[cPLI.ContextID]
\State PLI $\gets$ static + semi + cPLI.changing
\EndIf
\EndIf
\State return PLI
\end{algorithmic}
\end{algorithm}

The receiver algorithm is given in Algorithm~\ref{alg:temporal-compression-receiver} and functions as follows. Upon receiving a compressed PLI, the receiver determines the compression state. For uncompressed packets, the receiver stores the STATIC fields indexed by the UUID. For semi-compressed packets, the receiver retrieves the STATIC fields using the UUID and combines them with the received SEMI-STATIC fields to reconstruct the payload, and stores the static and semi-static fields together indexed by the Context-ID. For fully compressed packets, the receiver uses the received Context-ID to retrieve the full context (STATIC + SEMI-STATIC) and combines it with the CHANGING fields to recreate the original packet.

Errors may occur if a context-establishing packet is lost. If a receiver encounters a Context-ID or UUID that is not found in its local store, it transmits a Negative Acknowledgement (NACK) containing the missing identifiers. Upon receiving a NACK, the source re-broadcasts the full, uncompressed packet associated with that context. This re-initializes the temporal compression process. To prevent broadcast storms, the source utilizes a hold timer to aggregate multiple NACKs for the same context.







\subsection{Compression using Dynamic Compact Control Language (DCCL)}

Dynamic Compact Control Language (DCCL)~\cite{schneider2015dccl} is a language used for compressing message objects serialized by Google Protocol Buffers (GPB)~\cite{google_protobuf}. DCCL has found an extensive usage in extremely low capacity networks, such as underwater acoustic and satellite networks, where the throughput of a link may be 500~\textit{bps} or less. DCCL eliminates the overhead of field tags and byte-padding used by GPB to dissect a bytestream. Recent DCCL versions substitute the standard GPB encoder with more aggressive codecs that utilize domain-specific knowledge (i.e., value range, and accuracy) to achieve even higher compression rates, especially on \textit{integer} types.

In the context of a PLI message, each PLI field can be described with DCCL, defining the type of data, the range, and the accuracy of the value assigned. For example, a \textit{timestamp} field is defined as 64-bit integer carrying the current amount of elapsed milliseconds. With DCCL, we can define the range of desired time $\Delta T$ (e.g. from \textit{``now"} to \textit{``now + 1 day}"), and the accuracy of the value received $\delta$ (i.e. error tolerance), the amount of bits $N$ needed for encoding is:

\begin{equation}
\label{eq:DcclBitGain}
    N = \lceil \log_2 \left( \frac{\Delta T}{\delta} \right) \rceil
\end{equation}

With 2-ms error tolerance $\delta$, and the time-range $\Delta T$ equal to 1 day, the amount of bits $N$ needed for DCCL to encode a 64-bit \textit{Timestamp} field is 26 bits, translating into more than 50\% of compression gain.

We note that compression gains heavily depend on the data type (\textit{integer}, \textit{double}, \textit{bool}, \textit{byte}, etc.), and the possibility to tolerate a given range and accuracy of a value. The encryption-related and user-defined fields such as \textit{Encryption ID} or \textit{Callsign} carry arbitrary values, thus they cannot be effectively compressed using source encoding.

DCCL provides an efficient source encoding compression interface, which complements the \textit{Temporal Compression} algorithm described previously. It introduces an extra compression factor to CHANGING and SEMI-STATIC fields which couldn't be suppressed otherwise. For SQEEZ, DCCL provides a compression ratio of 0.82, 0.84 and 0.83 for static, semi-static and changing fields respectively.

\section{The Energy-Error Tradeoff}
\label{sec:tradeoff}

Consider a MANET with $N$ mobile nodes each moving 
with a velocity $v$. Assume each node is equipped with a location sharing application that network-wide broadcasts an PLI with default period $\defIntvl$. Let \Etx denote the energy per PLI transmission. We explore the tradeoff between the energy expended by the MANET nodes and the location error incurred, deriving expressions for each in the process. Our formulation applies to both the baseline location sharing as well as if SQEEZ is engaged. 

Let \Eav denote the energy consumption rate averaged over all nodes of the MANET. Let \lav denote the time- and node-averaged location error, i.e., the difference between a node's actual location and its inferred location based on PLI receptions (a more formal definition is given in section \ref{sec:performance-metrics}). Let $\tau$ denote the time between successive PLI packet originations, and let \lat denote the message latency. For convenience, all of the notation used is summarized in Table~\ref{tab:notation}.

Let $M$ be the number of nodes in the MANET that retransmit the update. If simple flooding is used for network-wide broadcast, then $M=N$. If an efficient flooding protocol such as ECHO~\cite{Ramanathan18} is used, then $M$ could be much smaller than $N$. Then

\begin{equation}
\label{eq:Eav}
    \Eav = \frac{M N \Etx}{\tau}
\end{equation}

\begin{table}[htbp]
\centering
\caption{Summary of Notation}
\label{tab:notation}
\begin{tabular}{lp{5.5cm}}
\toprule
\textbf{Symbol} & \textbf{Definition} \\
\midrule
$N$ & Number of mobile nodes in the MANET \\
$v$ & Node velocity \\
$\defIntvl$ & Default PLI broadcast period \\
\Etx & Energy per PLI transmission \\
\Eav & Average energy consumption rate \\
\lav & Time- and node-averaged location error \\
$\tau$ & Time between successive PLI originations \\
\lat & Message latency (average across nodes) \\
$M$ & Number of nodes that retransmit the update \\
$\Delta$ & Distance threshold \\
$B_{PLI}$ & Length of the PLI packet \\
$r$ & Bitrate \\
$P_{tx}$ & Default transmit power \\
$A_P$ & Amplification coefficient at power $P_{tx}$ \\
$I$ & Idle current \\
$V$ & Supply voltage \\
\bottomrule
\end{tabular}
\end{table}
For this analysis, we assume that the energy consumption is almost entirely due to transmissions. While this is not true in general, it is valid the context of the goTenna system where the transmit current draw is 250x of the receive/listen draw (see section~\ref{sec:evaluation} for more details). Incorporating receive energy into the analysis would require us to know the topology, which makes it unwieldy. 

We now consider the location error. Suppose an update was triggered at some time $t$. Then, in the interval ($t$, $t+\tau$) before the next update, the location error as perceived by a receiver $r$ varies from $\lat_r v$ at time $t$ to a maximum\footnote{If the node is stationary for some period of the time or changes directions, the error may be less then this maximum.} of $\lat_r v + \tau v$ at time $t+\tau$, where $\lat_r$ is the latency at receiver $r$. Let $\lat$ denote the average latency across nodes. Then, the time- and node-averaged location error 

\begin{equation}
\label{eq:lav}
    \lav = \frac{\tau v}{2} + \lat v
\end{equation}

Multiplying equation \ref{eq:Eav} and \ref{eq:lav}, we get

\begin{equation}
\label{eq:lavEav}
    \Eav\lav = v\frac{M N \Etx}{\tau}\left(\frac{\tau}{2} + \lat\right)
\end{equation}

In lightly loaded networks with a diameter of a few hops -- which is quite typical in operational scenarios -- the latency is in the order of a few 100 ms whereas $\defIntvl$ and hence $\tau$ is several 10's of seconds. Thus, to a first order approximation, we can assume $L \ll \tau$, and simplify equation \ref{eq:lavEav} as

\begin{equation}
\label{eq:lavEavApprox}
    \Eav\lav \approx v\frac{M N \Etx}{2}
\end{equation}

This captures the energy-error tradeoff. Specifically, it shows that for a given MANET scenario, protocol stack and PLI packet size (which determines $\Etx$), 
one cannot decrease the location error $\lav$ without incurring a corresponding increase in the energy expended $\Eav$, and vice versa. As the velocity increases, one can either maintain the same accuracy using higher energy consumption rate or maintain the same energy consumption while compromising on accuracy, but cannot do both. When the latency is negligible, the energy-error tradeoff is, remarkably, independent both of the time $\tau$ between updates as well as the latency. Equation~\ref{eq:lavEavApprox} is the basis, with suitable adaptation, for a novel metric \textit{error-penalized-energy} that we shall explore in section~\ref{sec:performance-metrics}.

We now proceed to express equations~\ref{eq:Eav} and \ref{eq:lav} in terms of network and configuration parameters alone by substituting for $\tau$. For the baseline case (no suppression), $\tau$ = $\defIntvl$. For suppression, let $\Delta$ be the distance threshold. 
A node takes $\frac{\Delta}{v}$ seconds to cross the distance threshold after which it is obliged to send an update at the next multiple of $\defIntvl$.  Specifically,

\begin{equation}
\label{eq:tau}
    \tau = \lceil(\frac{(\Delta/v)}{\defIntvl})\rceil \defIntvl 
\end{equation}

We note that for the baseline case of $\Delta$=0, $\tau$ instantiates to $\defIntvl$ in equation~\ref{eq:tau}, and so the equation captures both the baseline and SQEEZ.

Substituting equation~\ref{eq:tau} in equations \ref{eq:Eav} and \ref{eq:lav} respectively, we have the expressions:

\begin{equation}
\label{eq:EavFull}
    \Eav = \frac{M N \Etx}{\lceil(\frac{(\Delta/v)}{\defIntvl})\rceil \defIntvl}
\end{equation}

\begin{equation}
\label{eq:lavFull}
    \lav = \frac{\lceil(\frac{(\Delta/v)}{\defIntvl})\rceil \defIntvl  v}{2} + \lat v
\end{equation}

The above applies to both suppression with arbitrary $\Delta$ and the baseline which is essentially $\Delta=0$.

The \textit{continuous} version of suppression (see section~\ref{sec:sqeez-algorithms}), is equivalent to an very small $\defIntvl$ relative to $\Delta/v$ and therefore we can remove the "ceiling" operators\footnote{Note that the difference between $\lceil x/y \rceil$ and $x/y$ for any $x$, $y$ is at most 1 and therefore negligible if $x \gg y$.} from equations \ref{eq:EavFull} and \ref{eq:lavFull}. Thus, for the continuous version, we have

\begin{equation}
\label{eq:EavContinuous}
    \Eav = v \frac{M N \Etx}{\Delta}
\end{equation}

\begin{equation}
\label{eq:lavApprox}
    \lav = \frac{\Delta}{2} + \lat v
\end{equation}

The number of nodes $M$ that retransmit the PLI is a function of the broadcasting protocol and topology. If Flooding is used, then M = N irrespetive of the topology.  \Etx is a function of the packet size, data rate and hardware details. For the goTenna system modeled in the simulation, this is approximated by:

\begin{equation}
\label{eq:Etx}
\Etx = (\frac{B_{PLI}}{r})(\frac{P_{tx}}{A_P V} + I)V
\end{equation}

where $B_{PLI}$ is the length of the PLI packet, $r$ is the bitrate, $P$ is the default transmit power, $A_P$ is the amplification coefficient at power $P$, $I$ is the idle current and $V$ is the supply voltage.
\section{Evaluation}
\label{sec:evaluation}
In this section we discuss the evaluation of SQEEZ using a model of SQEEZ running over the Aspen Grove protocol stack~\cite{Ramanathan19}. We first define the performance metrics used for evaluation, followed by model details and simulation results.

\subsection{Performance Metrics}
\label{sec:performance-metrics}

The objective of SQEEZ is to provide energy-efficient location information dissemination with minimal compromise on location accuracy. Accordingly, it is important to measure the \textit{transmission energy} (with appropriate averaging) and the average \textit{location error}, which is the average difference between the \textit{actual} node locations and the \textit{believed} locations based on PLI packet receptions.   

As discussed in section~\ref{sec:tradeoff}, energy and error are locked in a tradeoff, that is, reducing error implies increasing energy. Comparing two algorithms based on energy or error alone is misleading -- for example, a protocol that consumes less energy than another could do so by greatly compromising on location error. Therefore, in addition to average energy and error we have developed a novel metric called \textit{Error Penalized Energy} (EPE), that combines both energy and error as a better way of comparison. 

We define each of these metrics below. In the below, $N$ is number of nodes and and $T$ is the deployment/simulation time over which the measurement is made.

\subsubsection{Average (Transmission) Energy Consumption} This is the average energy expended per minute per node for transmission purposes. It is computed by calculating the total energy expended by each transmission over the course of the simulation and normalizing it per minute and per node. For the goTenna Pro X, which is the radio on which our simulations are modeled, the default transmission power is 5W, at which the current draw is 2730 mA for transmission compared to only 40 mW for idle and reception, and therefore we use the transmission energy as a first order approximation for the total energy. The energy consumed by each individual transmission is given by equation~\ref{eq:Etx}, with the values of the dependent parameters in the equation as given in Table~\ref{tab:sim-parameters}. Let $K$ denote the total number of transmissions across the MANET and $E^i_{tx}$ denote the energy consumed for the $i^{th}$ transmission\footnote{Because of compression and other reasons, each packet may have a different length and hence consume different energy}. Then the Average Energy Consumption is calculated as:
\begin{equation}
\text{AEC} = \frac{ \sum_{i=1}^{K} E^i_{tx}}{N \times T}
\end{equation}

\subsubsection{Average Location Error}
We define the location error as the difference between the actual node locations and the believed locations based on PLI packet receptions. Specifically, if $L_{actual}(t)$ is the ground truth location at time $t$ and $L_{believed}(t)$ is the location estimated by the receiver based on the last received PLI, the error at time $t$ is the Euclidean distance between them. We define two averaging metrics for the location error, both commonly used in statistics. The \textit{Mean Absolute Error (MAE)} is defined as:

\begin{equation}
\text{MAE} = \frac{\sum_{t=0}^{T} \sum_{i=1}^{N} \text{dist}(L_{actual}^{(i)}(t), L_{believed}^{(i)}(t))}{N \times T}
\end{equation}

The \textit{Root Mean Square Error} is defined as: 
\begin{equation} 
\text{RMSE} = \sqrt{\frac{\sum_{t=0}^{T} \sum_{i=1}^{N} (\text{dist}(L_{actual}^{(i)}(t), L_{believed}^{(i)}(t)))^2}{N \times T}} 
\end{equation} 
While MAE provides a general sense of drift, the RMSE penalizes larger errors more heavily, which is critical for safety applications where large location discrepancies can be dangerous.

\subsubsection{Error Penalized Energy}
Comparing protocols solely on energy is often insufficient, as energy consumption can be trivially reduced by decreasing the update frequency, which negatively impacts accuracy. In section \ref{sec:tradeoff} we showed that reducing energy tends to increase location error, and therefore performance must be graded commensurate to the error increase. Accordingly, we introduce the Error-Penalized-Energy (EPE) metric. EPE ``normalizes" the energy consumption by the amount of error introduced relative to a baseline tolerance. It is defined as:
\begin{equation}
\text{EPE} = \text{AEC}\cdot \left( 1 + \left( \frac{Err_{av}}{Err_{base}} \right)^k \right)^{1/k}
\end{equation}
where AEC is the average energy consumed per minute per node, $Err_{av}$ is the average location error, $Err_{base}$ is the baseline error tolerance (e.g., 10m), and $k$ is a steepness factor (e.g., $k \approx 10$). For example, a protocol consuming 10 Joules with a 100m error has a higher EPE than one consuming 20 Joules with a 10m error, meaning that the latter is ``better" in spite of higher energy consumption, in line with intuition. The EPE is almost equal to AEC for $Err_{av} \leq Err_{base}$, and increases swiftly (linearly) for $Err_{av} > Err_{base}$, capturing intuitive notions.

\subsubsection{Packet Delivery Ratio}
The ratio between the number of PLIs received and the number of packets expected to be received. If S is the total number of PLIs originated and sent and R the total received, then 

\begin{equation}
PDR = \frac{R}{{(N-1) \times S}}
\end{equation}

\subsubsection{Scalability}
This addresses the question: how many nodes does the MANET scale to, for the given parameters? Following \cite{ramanathan2017symptotics} and adapting to our setting, we define it as the largest sized network for which the Packet Delivery Ratio is greater than 90\%. We find this by (binary) searching to find the point at which the PDR drops below 90\%. This metric only works for \textit{expandable} networks and scenarios, i.e., those based on a model~\cite{ramanathan2017symptotics} such as the Random Waypoint. It is not applicable to City Log.

\subsection{Simulation Model, Scenarios and Results}

\begin{table}[t]
\centering
\vspace{-0.2cm}
\caption{Simulation parameters.}
\vspace{-0.3cm}
\label{tab:sim-parameters}

\scriptsize
\setlength{\tabcolsep}{3pt}
\renewcommand{\arraystretch}{1.15}

\begin{tabularx}{\columnwidth}{@{}>{\raggedright\arraybackslash}p{0.47\columnwidth}>{\raggedright\arraybackslash}X@{}}
\toprule
\textbf{Simulation time per run, hours}     & 3                  \\
\textbf{Mobility Scenarios}                 & Citylog, Random Waypoint (RW) \\
\textbf{RW node velocity, mile/hour}             & [1-6] when swept over, 3 when not\\
\textbf{RW pause time, seconds}                      & [0-9000] when swept over, 600 when not\\
\textbf{MANET Size}                         & 9 (Citylog), 30 (RW) \\
\textbf{Laydown area, miles x miles}        & 3x3 default, 2x2 for scalability sims\\
\textbf{PLI period (Citylog), seconds}                & [5-30] when swept, 90 when not  \\
\textbf{PLI period (RW), seconds}                & [20-120] when swept, 90 when not  \\
\textbf{PLI Payload Size, bytes}            & 100 \\
\textbf{MANET Protocol Network / MAC }                        & ECHO\cite{Ramanathan18} / G-CSMA\\
\textbf{Radio bit rate, kbps}                     & 19.2\\
\textbf{Compression soft-state refresh, min}  & 120 \\
\textbf{Suppression soft-state refresh ($\refIntvl$), min} & 5 \\
\textbf{Suppression dist. threshold $\Delta$, meters} & 10 \\

\bottomrule
\end{tabularx}

\vspace{-0.4cm}
\end{table}


\begin{figure*}[ht]
\centering
\captionsetup[sub]{font=footnotesize,skip=1pt}

\begin{subfigure}[t]{0.48\textwidth}  
  \centering
  \includegraphics[width=\linewidth]{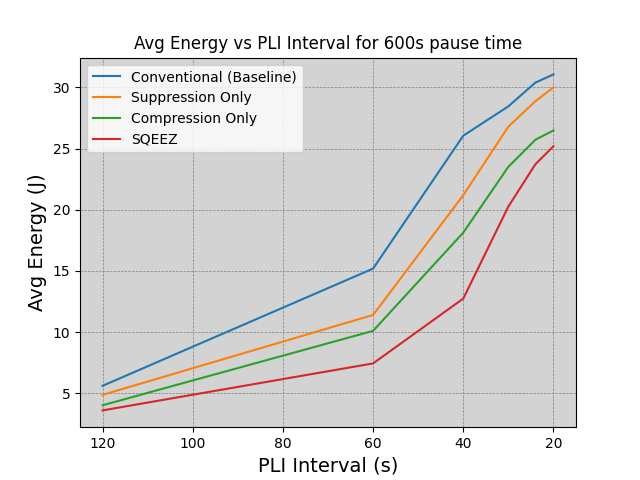}
  \subcaption{Average Energy Consumption, Joules}
  \label{fig:rwVsPLI-avEnergy}
\end{subfigure}\hfill
\begin{subfigure}[t]{0.48\textwidth}
  \centering
  \includegraphics[width=\linewidth]{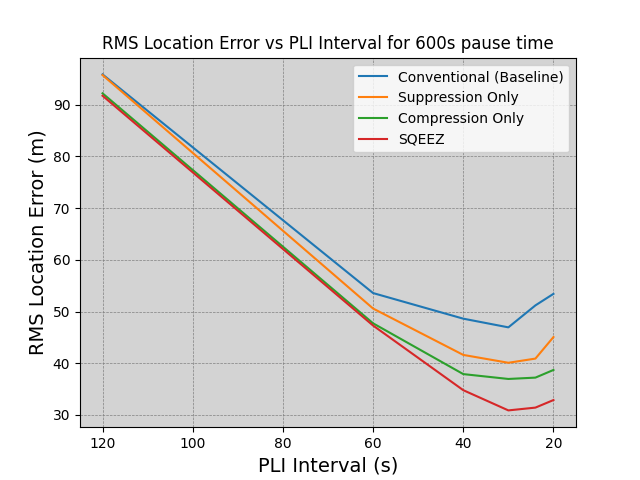}
  \subcaption{Location Error, meters}
  \label{fig:rwVsPLI-locErr}
\end{subfigure}

\vspace{0.3cm} 

\begin{subfigure}[t]{0.48\textwidth}
  \centering
  \includegraphics[width=\linewidth]{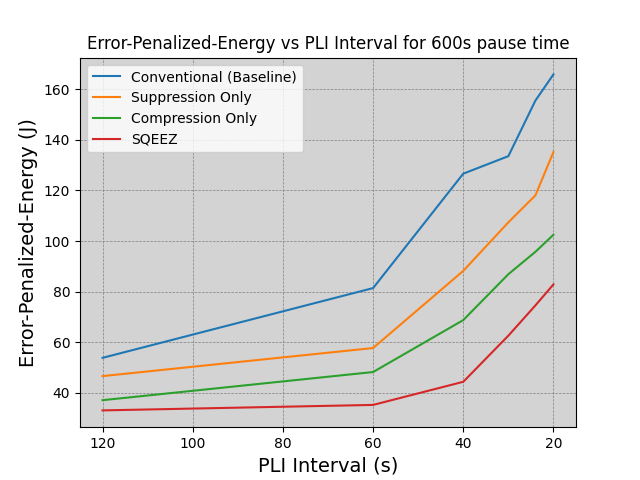}
  \subcaption{EPE, Joules}
  \label{fig:rwVsPLI-EPE}
\end{subfigure}\hfill
\begin{subfigure}[t]{0.48\textwidth}
  \centering
  \includegraphics[width=\linewidth]{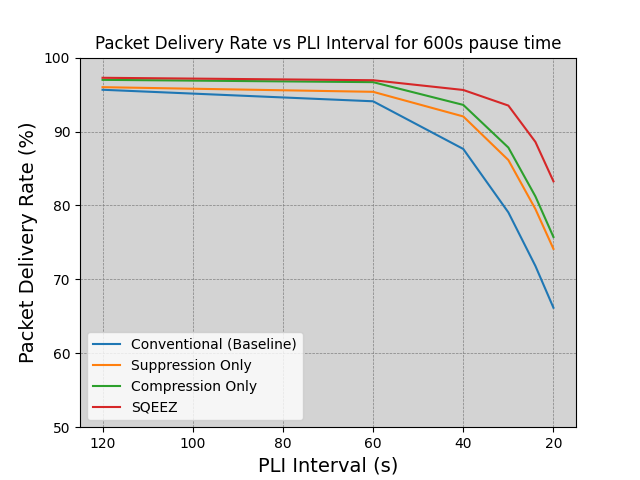}
  \subcaption{PDR, \%}
  \label{fig:rwVsPLI-PDR}
\end{subfigure}

\caption{30-node Random Waypoint results: Performance with varying PLI period}
\label{fig:rwVsPLI-results}
\end{figure*}

\begin{figure*}[ht]
\centering
\captionsetup[sub]{font=footnotesize,skip=1pt}

\begin{subfigure}[t]{0.48\textwidth}  
  \centering
  \includegraphics[width=\linewidth]{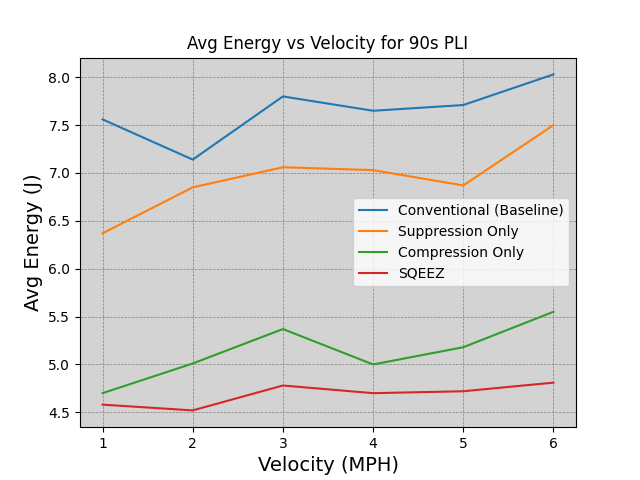}
  \subcaption{Average Energy Consumption, Joules}
  \label{fig:rwVsVel-avEnergy}
\end{subfigure}\hfill
\begin{subfigure}[t]{0.48\textwidth}
  \centering
  \includegraphics[width=\linewidth]{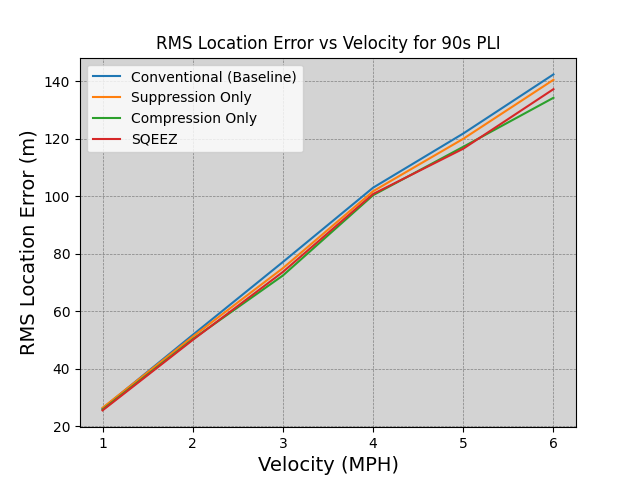}
  \subcaption{Location Error, meters}
  \label{fig:rwVsVel-locErr}
\end{subfigure}

\vspace{0.3cm} 

\begin{subfigure}[t]{0.48\textwidth}
  \centering
  \includegraphics[width=\linewidth]{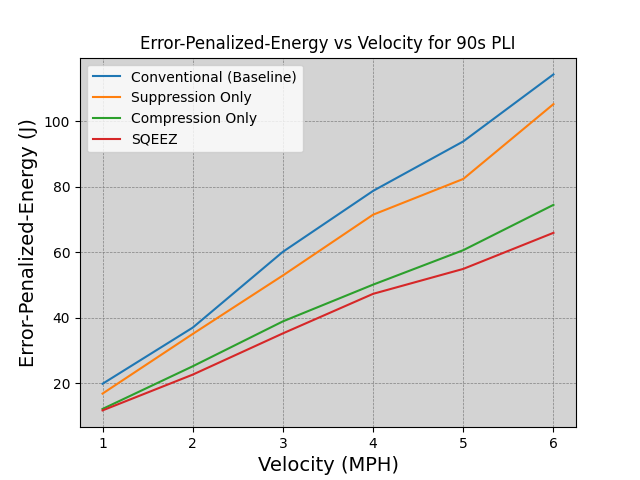}
  \subcaption{EPE, Joules}
  \label{fig:rwVsVel-EPE}
\end{subfigure}\hfill
\begin{subfigure}[t]{0.48\textwidth}
  \centering
  \includegraphics[width=\linewidth]{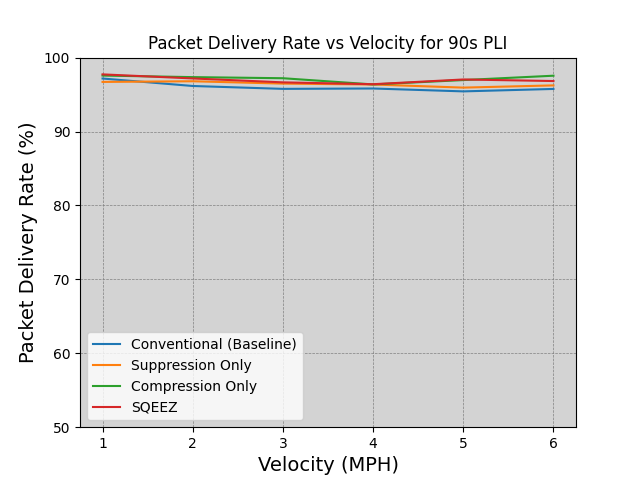}
  \subcaption{PDR, \%}
  \label{fig:rwVsVel-PDR}
\end{subfigure}

\caption{30-node Random Waypoint results: Performance with varying node velocity}
\label{fig:rwVsVel-results}
\end{figure*}

The simulations are based on a discrete-event model of SQEEZ running over the goTenna Aspen Grove stack using the goTenna Pro-X radios. The stack consists of the G-CSMA (``GoTenna CSMA") at the MAC layer, and the ECHO~\cite{Ramanathan18} broadcasting protocol at the network layer. The model is high-fidelity, incorporating packet sizes, queues, channel interference and packet capture, and has been calibrated against a testbed of goTenna radios.

We model two mobility scenarios: the \textit{random waypoint}~\cite{Navidi2004}, and \textit{City Log}. In random waypoint (RW), a node picks a random point in the deployed area and moves toward it at fixed velocity\footnote{Using a fixed velocity helps in alleviating the issue of convergence to the steady state~\cite{yoon2003random})}, pauses for a certain time at that location, and then repeats the process. Increasing the velocity or decreasing the pause time results in an increase in the extent of mobility. The ``City Log" mobility trace is an actual trace of 9 users navigating a city (Brooklyn, NY) with goTenna devices. The RW model allows for varying the size, velocity and pause time to provide a thorough study, while the City Log provides real-world validation. Table~\ref{tab:sim-parameters} contains a summary of the relevant parameters, some of which are fixed, and some are ``swept over" to plot dependence on it. 

The SQEEZ model follows the architecture in Figure~\ref{fig:squeez-arch}. For temporal compression, the simulation fixes the values of Encryption ID and Initialization Vector and sender callsign fields (see Table~\ref{tab:pli_fields}) to 3, 12 and 20 bytes respectively. Further, we mark semi-static fields whose persistence exceeds the refresh time of 2 hours as Static since they are essentially static for this duration.

We now discuss the evaluation results. Each plot has four curves: the no-SQEEZ conventional baseline, SQEEZ, SQEEZ with suppression only, and SQEEZ with compression only. As a visual aid, the reader can use ``blue" as "no-SQEEZ" and "red" as "SQEEZ" in all plots for a quick comparison.

Figure~\ref{fig:rwVsPLI-results} shows the performance for a 30-node Random Waypoint model as a function of PLI period. As expected, the average energy increases with decreasing PLI interval (more frequent updates), with SQEEZ providing up to a 2x improvement in energy efficiency (at 60s PLI) over the baseline. Compression by itself performs slightly better than suppression by itself, but this is specific to the particular mobility level (600s pause time). The location error shows an interesting dynamic, falling for most of the PLI range but increasing slightly toward the end. The decrease can be explained by the fact that more frequent updates reduce the location error; however, at low PLIs below 30s the network congestion/load is so high that there is significant packet loss which results in the location error going up. This is confirmed by the 
PDR curve that shows a steep drop off in delivery rate around 40s. The EPE curves show a similar trend as the energy curves, but with a slighlty larger separation since the location errors are slightly separated, penalizing the baseline more than SQEEZ. The EPE-efficiency of SQEEZ is 2.3x better than the baseline.

\begin{figure*}[ht]
\centering
\captionsetup[sub]{font=footnotesize,skip=1pt}

\begin{subfigure}[t]{0.48\textwidth}  
  \centering
  \includegraphics[width=\linewidth]{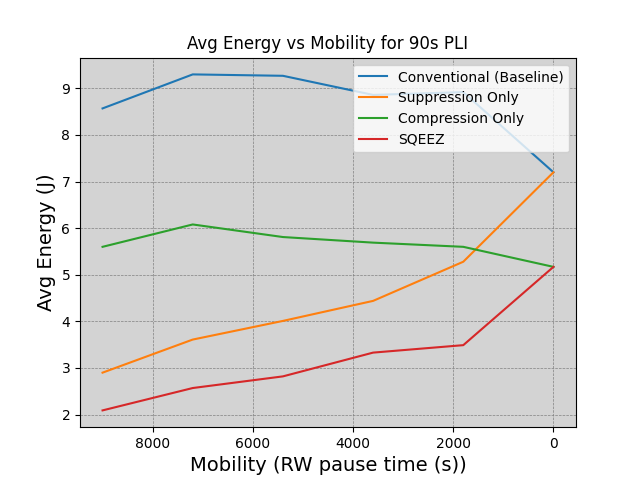}
  \subcaption{Average Energy Consumption, Joules}
  \label{fig:rwVsPause-avEnergy}
\end{subfigure}\hfill
\begin{subfigure}[t]{0.48\textwidth}
  \centering
  \includegraphics[width=\linewidth]{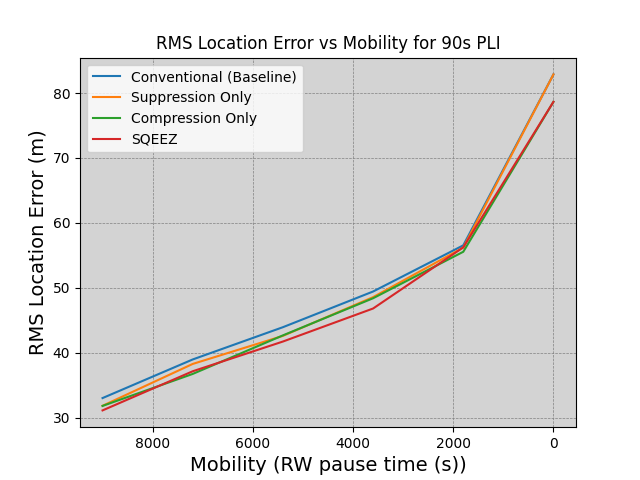}
  \subcaption{Location Error, meters}
  \label{fig:rwVsPause-locErr}
\end{subfigure}

\vspace{0.3cm} 

\begin{subfigure}[t]{0.48\textwidth}
  \centering
  \includegraphics[width=\linewidth]{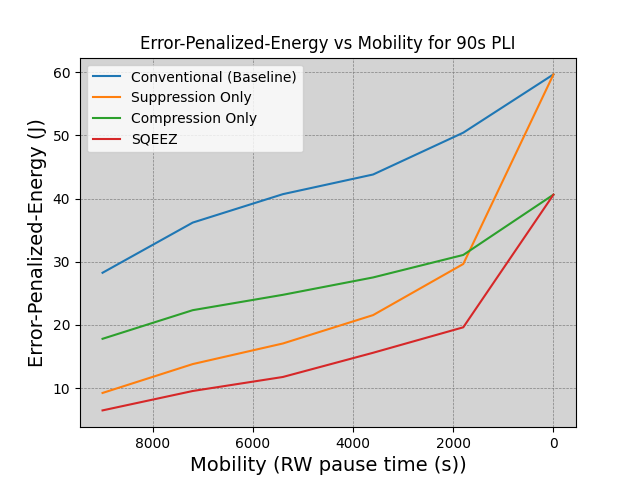}
  \subcaption{EPE, Joules}
  \label{fig:rwVsPause-EPE}
\end{subfigure}\hfill
\begin{subfigure}[t]{0.48\textwidth}
  \centering
  \includegraphics[width=\linewidth]{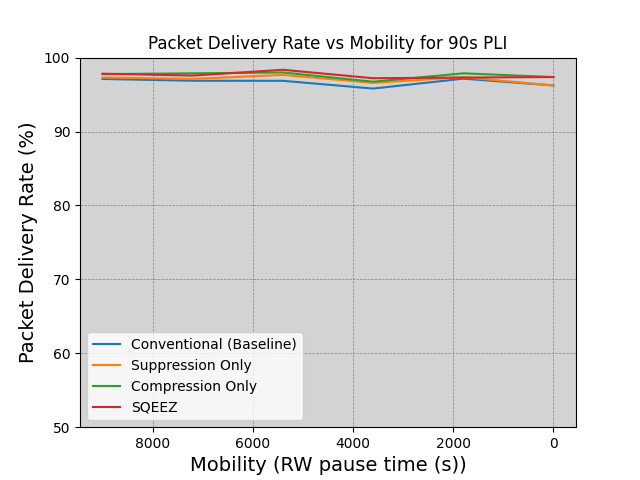}
  \subcaption{PDR, \%}
  \label{fig:rwVsPause-PDR}
\end{subfigure}

\caption{30-node Random Waypoint results: Performance with varying node wait time (pause time)}
\label{fig:rwVsPause-results}
\end{figure*}

Figure~\ref{fig:rwVsVel-results} shows the performance as a function of the node velocity. A key observation here is that even at the lowest velocity of 1 mph (0.44 m/s), a node will routinely cross the distance threshold of 10m betwee two successive PLI updates, and so suppression only happens if and when nodes change direction. The average transmit energy has little variation, which is not surprising since at these low velocities the energy is dominated by the default update interval. The location error increases with velocity as can be expected (see equation~\ref{eq:lav}), and because location is continuously changing and all variants have the same error since the PDR is high enough that few packets are dropped. The EPE trends upwards while maintaining the separation between SQEEZ variants due to the errors trending upward. Finally, the PDR is unremarkable at the low load of 90-second PLIs.

Figure ~\ref{fig:rwVsPause-results} shows the performance as the pause time decreases, i.e., effective mobility increases. Since suppression takes advantage of the stationarity of nodes to reduce transmissions, the energy consumption of ``suppression only" and SQEEZ increases with decreasing pause time, and at pause time 0, suppression does not yield any benefit. Consequently, the ``suppression only" touches the baseline curve and the SQEEZ curve touches the ``compression only" curve at pause time 0. At lower mobilities below pause time $\approx$ 1800s, suppression outperforms compression, and vice-versa at higher mobilities. SQEEZ combines the best of both and provides a 4.4x and 1.5x better EPE-efficiency at low- and high-mobility extremes respectively.
Figure \ref{fig:rwVsPause-locErr} shows that SQEEZ and its sub-variants do not compromise on the location error. The location error for all methods increase with less stationarity, since nodes are less likely to be at the location of the last update. The EPE curve steepens the curves from \ref{fig:rwVsPause-avEnergy} due to the error penalty which increases with mobility. Finally the PDR is near 100\% since the network is lightly loaded at 90s PLI.

\begin{figure*}[ht]
\centering
\captionsetup[sub]{font=footnotesize,skip=1pt}

\begin{subfigure}[t]{0.48\textwidth}  
  \centering
  \includegraphics[width=\linewidth]{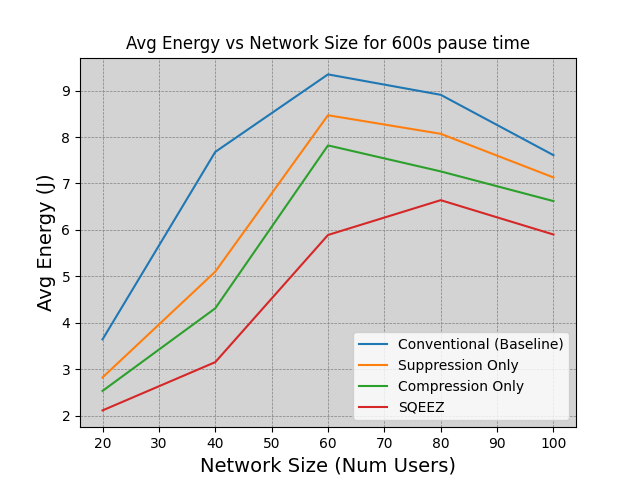}
  \subcaption{Average Energy Consumption, Joules}
  \label{fig:rwVsSize-avEnergy}
\end{subfigure}\hfill
\begin{subfigure}[t]{0.48\textwidth}
  \centering
  \includegraphics[width=\linewidth]{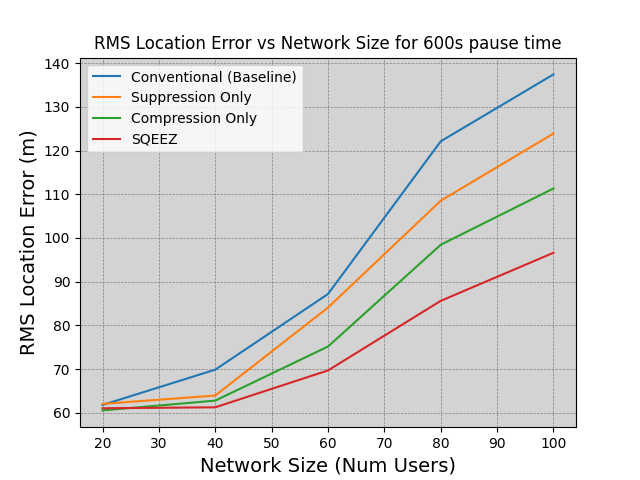}
  \subcaption{Location Error, meters}
  \label{fig:rwVsSize-locErr}
\end{subfigure}

\vspace{0.3cm} 

\begin{subfigure}[t]{0.48\textwidth}
  \centering
  \includegraphics[width=\linewidth]{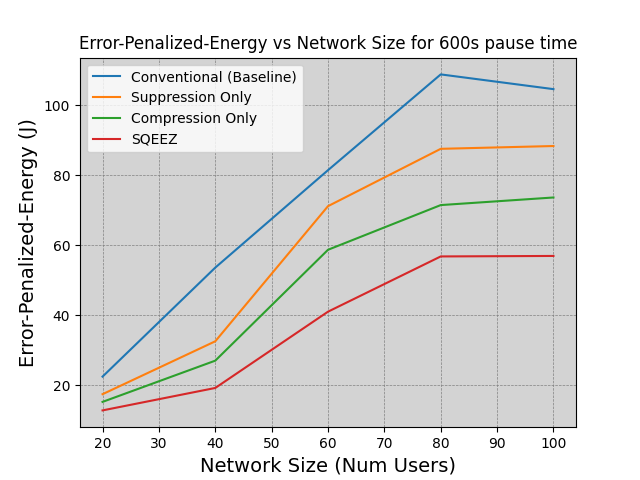}
  \subcaption{EPE, Joules}
  \label{fig:rwVsSize-EPE}
\end{subfigure}\hfill
\begin{subfigure}[t]{0.48\textwidth}
  \centering
  \includegraphics[width=\linewidth]{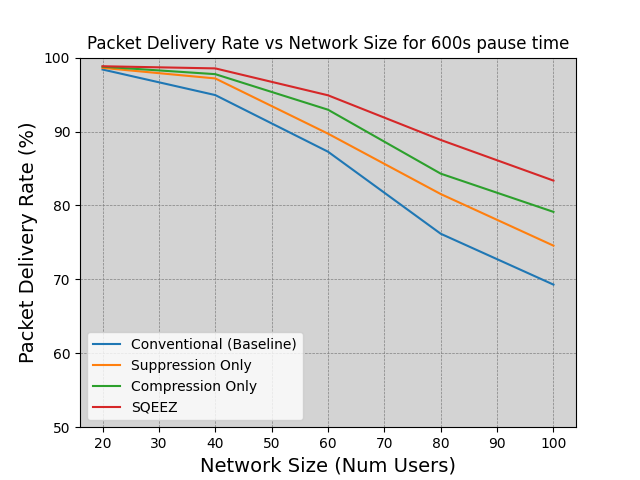}
  \subcaption{PDR, \%}
  \label{fig:rwVsSize-PDR}
\end{subfigure}

\caption{30-node Random Waypoint results: Performance with varying network size}
\label{fig:rwVsSize-results}
\end{figure*}

Figure \ref{fig:rwVsSize-results} shows the performance with varying network size (number of nodes). We first consider the PDR (Figure \ref{fig:rwVsSize-PDR}). Since each node originates broadcast PLIs, the load on the network can vary super-linearly with increasing size, which leads to rapid deterioration in the PDR. SQEEZ, as well as its sub-variants reduce the load somewhat, leading to correspondingly better PDRs. The location error (Figure \ref{fig:rwVsSize-locErr}) increases with size due to the higher loss rates, and once again, SQEEZ is able to do better by reducing the load. The average energy of all methods (Figure \ref{fig:rwVsSize-avEnergy}) falls after about 60 nodes, which may seem counter-intuitive, but can be explained by the falling PDR which ends up actually reducing the total number of packets. The trend is similar for EPE in Figure \ref{fig:rwVsSize-EPE}; the flattening of the curves at higher sizes can be explained by the higher location error which results in higher penalties in the EPE. In terms of relative performance, SQEEZ is not surprisingly the best, but it is interesting to note that ``compression only" outperforms ``suppression only" in terms of average energy, location error and EPE, while underperforming on PDR. At 40 nodes, SQEEZ provides a 2.75x improvement in EPE-efficiency over the baseline while having a better PDR.

\begin{figure*}[ht]
\centering
\captionsetup[sub]{font=footnotesize,skip=1pt}
\begin{subfigure}[t]{0.48\textwidth}
  \includegraphics[width=\linewidth]{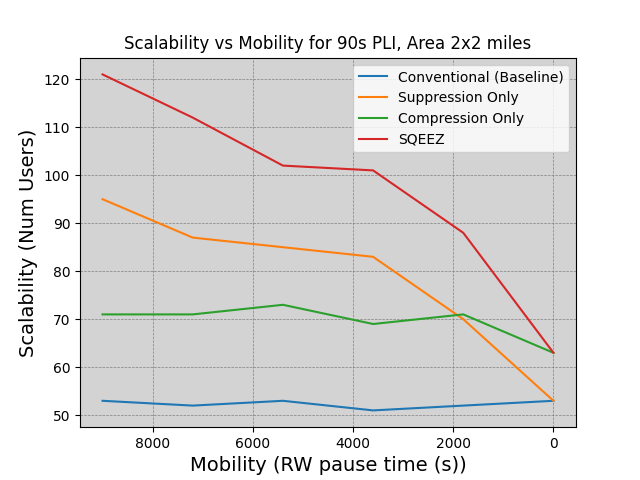}
  \subcaption{Fixed Area, Increasing Density}
  \label{fig:rw-scalability-FA}
\end{subfigure}
\begin{subfigure}[t]{0.48\textwidth}
  \includegraphics[width=\linewidth]{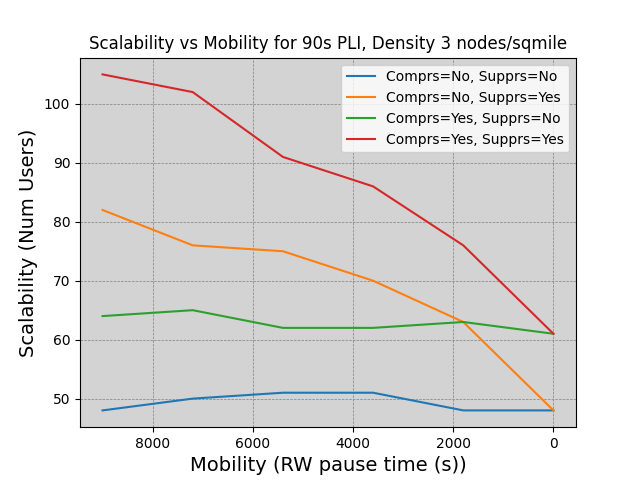}
  \subcaption{Fixed Density, Increasing Area}
  \label{fig:rw-scalability-FD}
\end{subfigure}
\vspace{-0.2cm}
\caption{\small \emph{Scalability Results, Fixed Area and Fixed Density}}
\label{fig:rw-scalability-results}
\vspace{-3mm}
\end{figure*}

Figure~\ref{fig:rw-scalability-results} shows the number of nodes to which the network scales with a $>$90\% PDR threshold. In 
Figure~\ref{fig:rw-scalability-FA}, the area of deployment was held constant as the network size increased, whereas in \ref{fig:rw-scalability-FD} the area increased proportionally to hold the density constant at 3 nodes/sq km. The results are more or less similar. At the lowest mobility, suppression is very effective, and provides a majority of the scalability benefits of SQEEZ. As mobility increases, fewer updates can be suppressed; at a pause time of 0 seconds, the scalability of the suppression-only scheme is same as the baseline. Compression is more immune to mobility except at high mobilities when the "Coordinate" field changes often, and forces the SEMI-STATIC fields to be included. This increases the packet size, which in turn increases the probability of packet collisions and reduces the PDR causing a corresponding reduction in scalability.

\begin{figure*}[ht]
\centering
\captionsetup[sub]{font=footnotesize,skip=1pt}

\begin{subfigure}[t]{0.48\textwidth}  
  \centering
  \includegraphics[width=\linewidth]{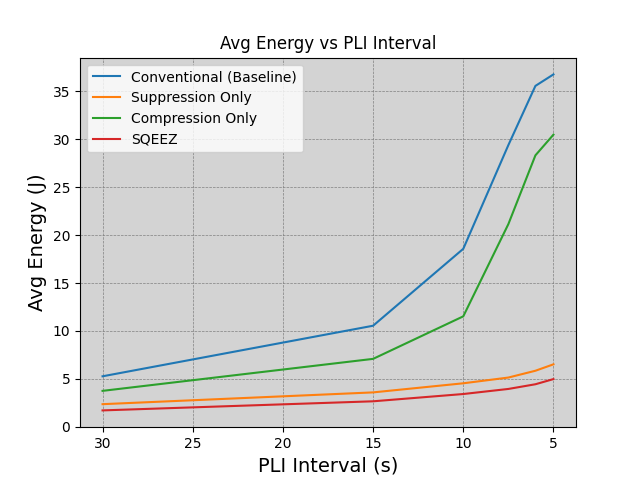}
  \subcaption{Average Energy Consumption, Joules}
  \label{fig:citylog-avEnergy}
\end{subfigure}\hfill
\begin{subfigure}[t]{0.48\textwidth}
  \centering
  \includegraphics[width=\linewidth]{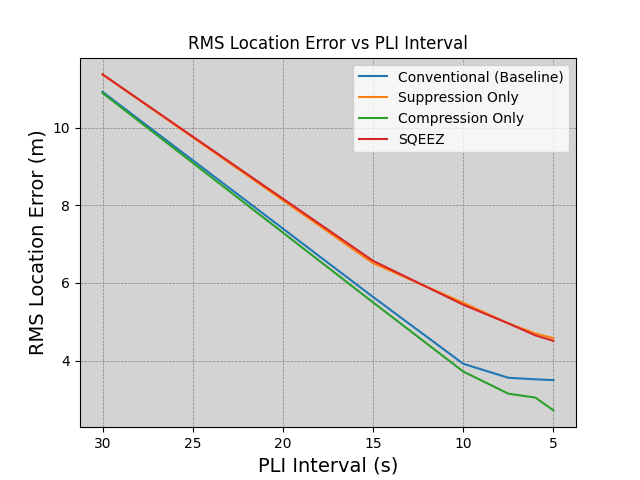}
  \subcaption{Location Error, meters}
  \label{fig:citylog-locErr}
\end{subfigure}

\vspace{0.3cm} 

\begin{subfigure}[t]{0.48\textwidth}
  \centering
  \includegraphics[width=\linewidth]{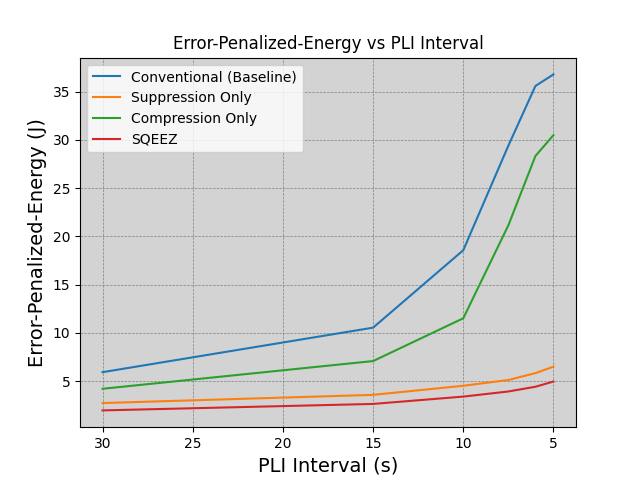}
  \subcaption{EPE, Joules}
  \label{fig:citylog-EPE}
\end{subfigure}\hfill
\begin{subfigure}[t]{0.48\textwidth}
  \centering
  \includegraphics[width=\linewidth]{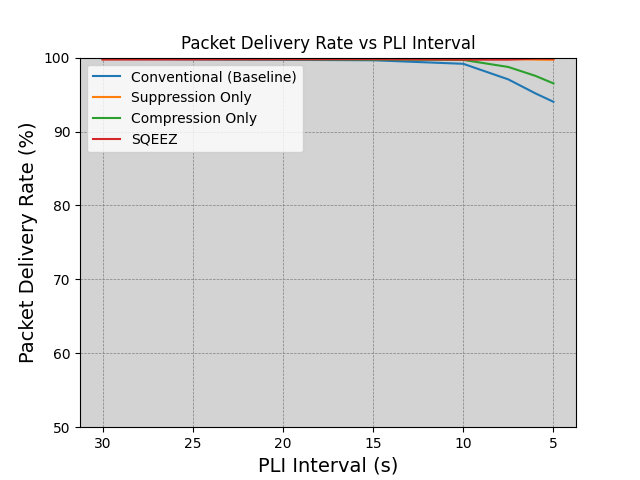}
  \subcaption{PDR, \%}
  \label{fig:citylog-PDR}
\end{subfigure}

\caption{Results using the Citylog trace, a real-world mobility trace}
\label{fig:citylog-results}
\end{figure*}

Figure~\ref{fig:citylog-results} shows the EPE for the 9 node City Log scenario as a function of the PLI interval. The PLI interval range has been lowered relative to the Random Waypoint experiments due to the smaller network size. The average energy (Figure~\ref{fig:citylog-avEnergy}) "suppression only" curve is very close to the SQEEZ curve (which has both suppression and compression), indicating that suppression contributes a majority of gains. This is because the size is small and mobility is low (walking speed with lots of pauses). Unlike in Figure~\ref{fig:rwVsPLI-locErr}, the location error does not go up at low PLI intervals. This is because, due to the small number of nodes, the packet loss remains low even at low PLI intervals and does not counteract the location error decrease. The EPE of SQEEZ is about 5 joules compared to 37 joules for the baseline, representing a 7.5x improvement in error-penalized-energy-efficiency. The PDR for all methods is near 100\% except at very high update rate as can be expected with a small network.

\section{Related Work}
\label{sec:related}

We discuss relevant work in three parts: suppression, temporal compression, and DCCL-compression.

Much of the existing published literature on location update suppression originates from geographic routing protocols. In the Distance Routing Effect Algorithm for Mobility (DREAM)~\cite{Basagni1998} protocol, the update frequency is a function of the node's mobility rate and the distance separating the sender and receiver. The Adaptive Position Update (APU) strategy~\cite{chen2012adaptive} reduces the frequency of beaconing by employing Mobility Prediction (MP) and On-Demand Learning (ODL) rules. 
Similarly, the AFB-GPSR protocol~\cite{Al-Essa2023} utilizes fuzzy logic to adaptively adjust beacon intervals based on speed and neighborhood density.  Unlike these methods which focus on routing-layer control messages and/or have a one-hop neighborhood scope, SQEEZ operates at the application layer for full PLI broadcast, and can work without predictable mobility. In~\cite{ye2010optimal}, the problem of Neighborhood Update (NU) and Location Server Update (LSU) for a generic location service is analyzed using a stochastic sequential decision framework, in particular using a Markov Decision Process. However, the focus is on theoretical properties.

Compression techniques in constrained wireless networks aim to reduce the payload size of the updates without loss. Robust Header Compression (ROHC)~\cite{Chen2012} is a well-established standard for reducing repetitive header fields in IP-based wireless links. Lossless compression algorithms have been proposed for data compression in wireless sensor networks, including Sensor LZW (S-LZW) \cite{Sadler2006} and Lossless Entropy
Compression (LEC) \cite{Marcelloni2008}. Temporal compression is also the basis for video encoding; however, the problems and solutions tend to be quite different from PLI compression. We note that the term "temporal compression" has also been used with other meanings such as in \cite{Klus2021} where the intent is to compress time-series data using patterns -- a completely different problem.

Dynamic Compact Control Language (DCCL)~\cite{schneider2015dccl} is a specialized interface description language (IDL) and serialization library designed to address communication challenges in extremely low throughput networks. DCCL provides a mapping interface between high-level object-based messages, such as Google Protocol Buffers (GPB)~\cite{google_protobuf}, and translates it into a raw bitstream with bounded field definitions, compressed using source encoding algorithms. Due to very high compression efficiency, especially on 32/64-bit integer types, it is widely used for acoustic communications in underwater networks~\cite{schneider2010dynamic,alves2011low}. We are not aware of any usage for PLI compression as in SQEEZ.

ATAK and its variants work in tandem with these network constraints created by user deployments and mission. The rise of electronic warfare and near pear capabilities has created a demand on efficient usage of on air time for any transmission for mission critical information of any MANET network type. The need for Blue Force tracking and asset location is one of the most critical information points in the success of any mission in these types of conditions.

SQEEZ is unique in its holistic integration of application-layer suppression with a dual-anchor temporal compression scheme. 
Furthermore, by combining these with bit-level DCCL marshalling, SQEEZ achieves a degree of information density that enables situational awareness applications like ATAK~\cite{Usbeck2015} to scale to mobilities and network sizes previously infeasible on narrow-band tactical links.

\section{Conclusions}
\label{sec:conclusions}

We described SQEEZ -- a mechanism that combines adaptive suppression, temporal and inline compression of Position Location Update (PLI) packets to reduce the traffic load on a MANET. SQEEZ suppresses PLI originations within a threshold distance of the previously updated location; compression uses a dual-anchor scheme to exploit the temporal redundancy in ATAK-related PLI messages as well as compression of the resulting message using DCCL. The SQEEZ architecture allows it to coexist with a baseline system.

We quantified the tradeoffs between the (average) transmit energy used and location error, and derived expressions for each. A key observation is that the product of the average location error and transmit energy for a given MANET deployment with a given average node velocity is a constant; thus, one can either reduce location error or transmit energy but not both. To properly account for the loss in location accuracy, we defined a novel metric \textit{Error-Penalized-Energy (EPE)} in addition to other conventional metrics. Our simulations using the Random Waypoint model showed that SQEEZ improves the EPE-efficiency by up to a factor of 1.3x-5.6x depending on the amount of stationarity of nodes, and improves scalability by up to a factor of 2.3x. We observed that compression contributed a larger portion of SQEEZ's gains at high mobilities, whereas suppression contributed more at lower ones. In a real-world ``CityLog" mobility trace, the gains were as much as 7.5x.

We conclude that SQEEZ can be effective in significantly reducing the load and hence the energy consumption, as well as improving the scalability of a location-sharing system such as ATAK over a mesh/ad hoc network. Our study brings out the importance of tracking location error and of using EPE in addition to raw energy consumption.

\bibliographystyle{IEEEtran}
\bibliography{ref}

\end{document}